\documentclass[%
 aip,
 amsmath,amssymb,
 reprint,%
]{revtex4-1}

\usepackage{graphicx}% Include figure files
\usepackage{dcolumn}% Align table columns on decimal point
\usepackage{bm}% bold math
\usepackage[utf8]{inputenc}
\usepackage[T1]{fontenc}
\usepackage{mathptmx}
\usepackage{etoolbox}
\usepackage{bigints}
\usepackage[normalem]{ulem}
\usepackage[dvipsnames]{xcolor}

\makeatletter
\def\@email#1#2{%
 \endgroup
 \patchcmd{\titleblock@produce}
  {\frontmatter@RRAPformat}
  {\frontmatter@RRAPformat{\produce@RRAP{*#1\href{mailto:#2}{#2}}}\frontmatter@RRAPformat}
  {}{}
}%
\makeatother
\begin{document}

\preprint{AIP/123-QED}

\title[THF hydrate - water interfacial free energy]{Simulation of the THF hydrate - water interfacial free energy from computer simulation}
% Force line breaks with \\

\author{Miguel J. Torrej\'on}
\affiliation{Laboratorio de Simulaci\'on Molecular y Qu\'imica Computacional, CIQSO-Centro de Investigaci\'on en Qu\'imica Sostenible and Departamento de Ciencias Integradas, Universidad de Huelva, 21006 Huelva Spain}

\author{Crist\'obal Romero-Guzm\'an}
\affiliation{Laboratorio de Simulaci\'on Molecular y Qu\'imica Computacional, CIQSO-Centro de Investigaci\'on en Qu\'imica Sostenible and Departamento de Ciencias Integradas, Universidad de Huelva, 21006 Huelva Spain}

\author{Manuel M. Piñeiro}
\affiliation{Departamento de Física Aplicada, Universidade de Vigo, 36310 Vigo, Spain}
\author{Felipe J. Blas}
\affiliation{Laboratorio de Simulaci\'on Molecular y Qu\'imica Computacional, CIQSO-Centro de Investigaci\'on en Qu\'imica Sostenible and Departamento de Ciencias Integradas, Universidad de Huelva, 21006 Huelva Spain}

\author{Jes\'us Algaba$^*$}
\affiliation{Laboratorio de Simulaci\'on Molecular y Qu\'imica Computacional, CIQSO-Centro de Investigaci\'on en Qu\'imica Sostenible and Departamento de Ciencias Integradas, Universidad de Huelva, 21006 Huelva Spain}
\email{jesus.algaba@die.uhu.es}

\begin{abstract}

In this work, the tetrahydrofuran (THF) hydrate-water interfacial free energy is determined at $500\,\text{bar}$, at one point of the univariant two-phase coexistence line of the THF hydrate, by molecular dynamics simulation. The Mold Integration-Host methodology, an extension of the original Mold Integration technique to deal with hydrate-fluid interfaces, is used to calculate the interfacial energy. Water is described using the well-known TIP4P/Ice model and THF is described using a rigid version of the TraPPE model. We have recently used the combination of these two models to accurately describe the univariant two-phase dissociation line of the THF hydrate, in a wide range
of pressures, from computer simulation [J.~Chem.~Phys.~\textbf{160}, 164718 (2024)]. The THF hydrate-water interfacial free energy predicted in this work is compared with the only experimental data available in the literature. The value obtained, $27(2)\,\text{mJ/m}^{2}$, is in excellent agreement with the experimental data taken from the literature, $24(8)\,\text{mJ/m}^{2}$. To the best of our knowledge, this is the first time that the THF hydrate-water interfacial free energy is predicted from computer simulation. This work confirms that the Mold Integration technique can be used with confidence to predict solid-fluid interfaces of complex structures, including hydrates that exhibit sI and sII crystallographic structures.

\end{abstract}

\maketitle

%\begin{quotation}
%The ``lead paragraph'' is encapsulated with the \LaTeX\ 
%\verb+quotation+ environment and is formatted as a single paragraph before the first section heading. 
%(The \verb+quotation+ environment reverts to its usual meaning after the first sectioning command.) 
%Note that numbered references are allowed in the lead paragraph.
%
%The lead paragraph will only be found in an article being prepared for the journal \textit{Chaos}.
%\end{quotation}

\section{Introduction}

 Clathrate hydrates are crystalline inclusion compounds in which small molecules such as methane (CH$_4$), carbon dioxide (CO$_2$), hydrogen (H$_2$), or nitrogen (N$_2$) are encapsulated in the voids left by a network of hydrogen-bonded molecules. When the network of hydrogen-bonded molecules is built with water molecules (H$_2$O), clathrate hydrates are simply called hydrates.~\cite{Sloan2008a,Ripmeester2022a}  Hydrates have important applications due to the capability of these compounds to capture CO$_2$,~\cite{ma2016review,dashti2015recent,cannone2021review,duc2007co2,choi2022effective,lee2014quantitative} to store H$_2$ efficiently and safely,~\cite{veluswamy2014hydrogen,Lee2005a} to recover N$_2$ from industrial emissions,~\cite{Yi2019,hassanpouryouzband2018co2} and they are also interesting from an energetic point of view since there is more CH$_4$ as hydrates in the nature than in conventional fossil fuel reservoirs.~\cite{lee2001methane,ruppel2017interaction} Hydrates are formed under high pressures and low temperatures, however, the stability of these compounds can be dramatically improved by the use of additives.~\cite{Sloan2008a,Li2012a} In particular, thermodynamic promoter additives are able to reduce the pressure and/or increase the temperature at which hydrates can crystallize. 

Tetrahydrofuran (THF), a five-member cycle ether,  has been widely used as a hydrate promoter.  THF can enhance the stability of hydrates by lowering drastically the pressure at which they are stable.~\cite{Florusse2004a,Lee2005a,Strobel2007a,veluswamy2014hydrogen} THF occupies the large cages (5$^{12}$6$^4$) of the hydrate structure, while the small cages (5$^{12}$) can be filled with other small guest molecules. A unique characteristic of THF is its ability to form stable hydrates by itself. THF hydrate is stable at temperatures below 277 K and atmospheric pressure,~\cite{Makino2005a} which is considered mild conditions compared to the conditions required for most hydrates to be stable.~\cite{Sloan2008a}

The THF hydrate shows a univariant two-phase coexistence line in which the composition of the aqueous solution phase remains constant and equal to the composition of the hydrate (17 H$_2$O : 1 THF).~\cite{Asadi2019a,Sun2017a,Suzuki2011a,Sabase2009a} It means that this hydrate is formed when THF and water are mixed at the stoichiometric ratio found in the THF hydrate. This is possible because, at the temperature and pressure conditions at which the hydrate can be formed, THF and water are completely miscible at any ratio. Although most experiments are performed at stoichiometric conditions, several works have been devoted to studying how the THF hydrate is formed at higher and lower THF concentrations in the aqueous solution phase, due to its interest as a thermodynamic additive mixed with other guests of interest.~\cite{Strauch2018a,Ganji2006a,Andersson1996a,Chong2016a,Kumar2010a} There is a range of THF composition in the aqueous solution phase (from $5.0$ to $82.7\,\text{wt}\%$) in which the formation of the THF hydrate is possible.~\cite{Strauch2018a} However, Liu \emph{et al.}~\cite{Liu2022a} demonstrated that even if the formation of the THF hydrate from an aqueous solution with a different composition than the THF hydrate is possible, the amount of THF in the hydrate remains invariable, and the large cages of the sII hydrate structure are fully occupied by a molecule of THF. As a consequence of this, when the hydrate crystallizes from an aqueous solution with a lower THF composition than the stoichiometric one, the hydrate grows until the aqueous phase reaches a THF concentration of $6.5\,\text{wt}\%$, approximately.~\cite{Liu2022a} However, if THF composition in the aqueous phase is higher than the stoichiometric one, the hydrate grows until the aqueous phase reaches a THF concentration of $44.0\,\text{wt}\%$, approximately.~\cite{Liu2022a} Also, it is important to remark that even if the THF hydrate can crystallize from a non-stoichiometric aqueous solution, the kinetic is affected, and the crystallization takes place slower than when the crystallization occurs from a stoichiometric aqueous solution.~\cite{Liu2022a} The results obtained by these authors provide insightful information about the thermodynamics and the kinetics of the THF hydrate at atmospheric pressure. 

Accurate knowledge of the thermodynamic conditions of stabilities of the THF hydrate is essential in order to study the nucleation of hydrates of interest when the THF is used as an additive. Another property of great interest is the interfacial free energy between the hydrate structure and the aqueous solution phase. The interfacial free energy has a key role in hydrate growth and nucleation.~\cite{Aman2016a,Sarupria2011a,Sarupria2012a,Zhan2018a,Wang2022a,Wang2023a} From the interfacial free energy value, it is possible to calculate simply the hydrate-liquid interfacial tension, $\gamma_{hw}$. As far as the authors know, there is only one value of $\gamma_{hw}$ for the THF hydrate reported in the literature.~\cite{Zakrzewski1993a} Although hydrates have been widely studied, there is still a poor understanding of the nucleation and growth mechanism from a molecular point of view.~\cite{Kashchiev2002a,Kashchiev2002b,Kashchiev2003a,Sarupria2011a,Walsh2011a,Sarupria2012a,Barnes2014a,Barnes2014b,Yuhara2015a,Grabowska2022b,Algaba2023a} In this context, some of us have recently extended the original Mold Integration (MI)  method~\cite{Espinosa2014a,Espinosa2016a} to hydrates.~\cite{Algaba2022b,Zeron2022a,Romero-Guzman2023a} The MI methodology provides an accurate and non-expensive method for solid-fluid interfacial free energy determination. In the case of the CO$_2$ hydrate, the results obtained by the extended versions of the MI method provide an excellent agreement between simulation results and the only two experimental values reported in the literature.  

\begin{figure}
%\hspace*{-0.7cm}
\centering
\includegraphics[width=\columnwidth]{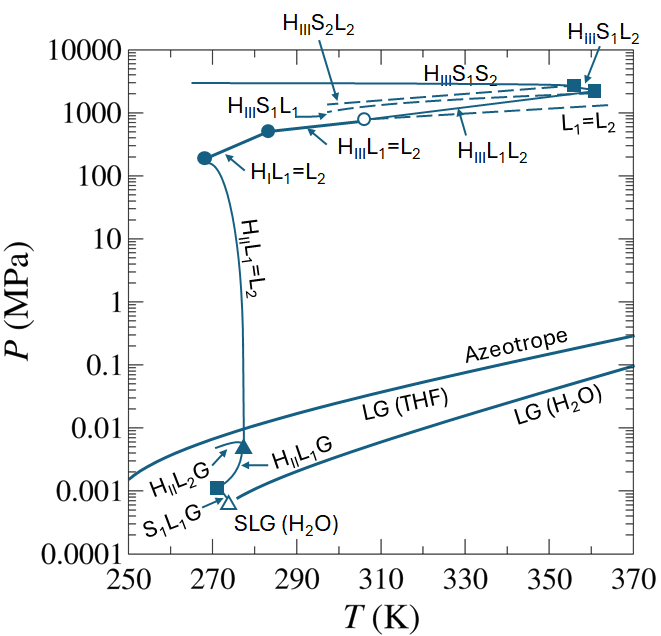}
\vspace*{-0.0cm}
\caption{Schematic representation of the experimental THF hydrate--water phase diagram.~\cite{Makino2005a} H$_{\text{I}}$, H$_{\text{II}}$, and H$_{\text{III}}$ stand for the sI, sII, and sIII (or sH) THF hydrate structures respectively, S$_{\text{I}}$ and S$_{\text{II}}$, for solid water (ice VII) and THF phase respectively,  L$_{\text{I}}$ and L$_{\text{II}}$ for an aqueous phase with a THF composition below and above the stoichiometric one respectively, and G for a gas phase. The meaning of each equilibrium curve is represented in the picture. The symbols represent the thermodynamic conditions at which two equilibrium curves cross and the phases involved are the sum of the phases of both equilibrium curves. Finally, the pure Liquid-Gas (LG) equilibrium curves for pure water and THF are also represented.}
\label{phase_diagram}
\end{figure}

To apply the MI method for solid-fluid interfacial free energy determination, it is necessary to have an accurate description of the dissociation conditions of the hydrate of interest since this method can be only used when the solid and fluid phases are in equilibrium. Very recently, some of us have studied the univariant two-phase coexistence line of the
THF hydrate from computer simulation.~\cite{Algaba2024d} In the seminar work of Algaba \emph{et al.},~\cite{Algaba2024d} the TIP4P/Ice model~\cite{Abascal2005b,Conde2017a} for water molecules and a rigid version TraPPE-UA THF~\cite{Keasler2012a,Garrido2016a,Algaba2018a,Algaba2019a} were used to describe the univariant two-phase coexistence line of the
THF hydrate, finding an excellent agreement between simulation and experimental results.  Notice that in our seminar work,~\cite{Algaba2024d} we focused on the THF hydrate phase diagram from $10$ to $100\,\text{MPa}$. This is only a portion of its global phase diagram that exhibits a much more complex behavior (see Fig.~\ref{phase_diagram}). At low pressures, from $1.1$ to $4.9\,\text{KPa}$, the THF hydrate shows two characteristic three-phase equilibrium (hydrate-aqueous-gas) curves. In the first one (H$_{\text{II}}$L$_1$G), the amount of THF in the aqueous phase is below the stoichiometric composition in the sII THF hydrate phase (1~THF:17~H$_2$O), while in the second one (H$_{\text{II}}$L$_1$G), the amount of THF in the aqueous phase is above the stoichiometric composition. 
H$_{\text{II}}$ denotes the hydrate phase (sII structure), L the water-rich liquid phase, and G the gas phase. At higher pressures, from $0.049$ to $200\,\text{MPa}$, the THF phase diagram exhibits a univariant hydrate–aqueous solution two-phase coexistence curve where the aqueous solution exactly has the stoichiometric composition of the hydrate. Above 200 MPa, the THF hydrate phase diagram becomes even more complex involving several hydrate phase transitions (from sII to sI and from sI to sIII). Besides, above 200 MPa, the solid phases of water (as ice) and THF are also involved in the THF-water-hydrate phase equilibria.

The main goal of this work, which has to be considered the natural continuation of the previous one, is to use the same molecular models to describe the interfacial free energy between the THF hydrate and a stoichiometric aqueous solution phase following the MI methodology. It is important to notice that the MI methodology is independent of the solid structure.  It has been implemented to calculate the interfacial free energy of several pure systems.~\cite{Espinosa2014a,Espinosa2016a,Espinosa2015a} The two extensions of this methodology proposed by some of us in previous works, the MI-Host~\cite{Algaba2022b,Romero-Guzman2023a} (MI-H) and the MI-Guest\cite{Zeron2022a} (MI-G), allow us to extend the MI technique to systems formed from two or more components as in the case of hydrate systems. In this work, we employ straightforwardly the extended version of the MI methodology (MI-H) to the THF hydrate case. The MI-H methodology has been employed by some of us to determine the CO$_2$ hydrate-water interfacial tension.~\cite{Algaba2022b,Romero-Guzman2023a} The fact that the CO$_2$ (sI) and the THF (sII) hydrates exhibit different crystallographic structures does not affect the implementation of the MI-H methodology as far as the mold resembles the corresponding crystallographic structure.

The organization of this paper is as follows: In Sec. II, we describe the extended version of the MI method used in this work. Simulation details are described in Sec. III. The results obtained in this work are discussed in detail in Sec. IV. Finally, conclusions are presented in Sec. V.

\section{Methodology}

This work estimates the THF hydrate-water interfacial free energy via the Mold Integration-Host method (MI-H).~\cite{Algaba2022b,Romero-Guzman2023a} Following this methodology, a number of potential wells ($N_{w}$) are placed in the middle of an aqueous phase at the crystallographic equilibrium positions of the oxygen atoms of water molecules in the THF hydrate sII crystallographic structure. The potential wells are then progressively activated from 0 to their maximum energy, denoted as $\varepsilon_m$, inducing the formation of a crystal slab, with the consequent creation of two THF hydrate-water interfaces. At this point, it is important to remark that the composition of the aqueous phase remains constant and equal to the composition of the THF hydrate since this is one of the peculiarities of this system, the THF hydrate presents an invariant composition (17 H$_2$O~:~1 THF) along its dissociation line.~\cite{Algaba2024d}

The free energy difference between the bulk aqueous phase and the aqueous phase with the hydrate slab, at coexistence conditions, can be obtained via thermodynamic integration, as originally described by Espinosa~\emph{et al.}~\cite{Espinosa2014a,Espinosa2016a} This allows to create a thermodynamic path that connects the aqueous fluid system whose wells are deactivated ($\varepsilon = 0$) with the aqueous fluid system whose wells are at their maximum energy value ($\varepsilon = \varepsilon_m$) and in which a thin slab of solid THF hydrate is created. The free energy difference between the aqueous system with the mold completely activated and deactivated is the sum of two contributions, the interaction between the mold and the system ($N_{w}\epsilon_{m}$), and the interfacial free energy, $\Delta G^{hw}$, due to the formation of the THF hydrate-water interface. After subtracting the mold-system contribution, the reversible work required to form a thin hydrate slab because of the formation of two THF hydrate-water interfaces is given by,

\begin{equation}
\Delta G^{hw}=N_{w}\epsilon_{m}-\int_{0}^{\varepsilon_{m}} \langle N_{fw}(\varepsilon) \rangle _{NP_{z}\mathcal{A}T}d\varepsilon \,
\label{eq:thermo-int-1}
\end{equation}

\noindent
where $\langle N_{fw}(\varepsilon)\rangle _{NP_{z}\mathcal{A}T}$ is the averaged number of filled wells at each $\varepsilon$ value along the thermodynamic integration at coexistence conditions. The angular bracket in Eq.~\eqref{eq:thermo-int-1} denotes a configurational average in the isothermal-isobaric or ${NP_{z}\mathcal{A}T}$ ensemble. This ensures that $\Delta G^{hw}$ is calculated at coexistence conditions $P_{z}$ and $T$. Here $P_{z}$ is equal to the macroscopic equilibrium pressure of the system. Note that we use the anisotropic ${NP_{z}\mathcal{A}T}$ ensemble in which $L_{z}$, the size of the simulation box along the $z$-axis, is allowed to fluctuate. According to this, the interfacial area $\mathcal{A}=L_{x}\times L_{y}$, with $L_{x}$ and $L_{y}$ the size of the simulation box along the $x$- and $y$-axis directions, is kept constant and equal to the equilibrium values of the THF hydrate phase at the same thermodynamic conditions. 

The interfacial free energy obtained through the MI-H method is independent of the mold and its properties. However, three parameters require careful selection for an accurate application of the method. First, an appropriate value of $\varepsilon_m$ needs to be selected. To subtract from the total free energy difference the mold-system contribution, once the mold is completely activated, the oxygen atoms from the water molecules cannot escape from it and only one atom of oxygen must occupy each well. On the one hand, if $\varepsilon_m$ is too low, the mold will not be able to retain the oxygen atoms of the water molecules, allowing them to escape. On the other hand, if $\varepsilon_m$ is too high, more than one oxygen from different water molecules could be retained in the well. Second, the number of wells, $N_{w}$, must be large enough to provide a representative picture of the crystalline structure of the solid phase. Since, following Eq.~\ref{eq:thermo-int-1}, the mold-system energy contribution ($N_{w}\epsilon_{m}$) to the free energy is subtracted, the final result of the interfacial free energy is independent of the values of $\varepsilon_m$ and $N_w$ as has been demonstrated by some of us in previous works.~\cite{Zeron2022a,Romero-Guzman2023a} And third, the interfacial free energy depends on the potential well length, $r_{w}$, which determines the region where the mold attracts and retains the oxygen of the water molecules, and its contribution can affect whether the energy barrier can be overcome or not. If $r_{w}$ is too small, the molecules trapped inside are fixed at the ideal crystallographic positions of the induced slab. Hence, the energy given from the mold to the system is too high, the fluid phase becomes unstable, and the system crystallizes as soon as the simulation begins (scenario I). However, if $r_{w}$ is too high, the molecules inside the potential wells can move with more freedom and the slab is unlikely to be created. As a result of the increase in freedom, the system cannot overcome the energetic barrier immediately and the system does not crystallize upon starting the simulation. In this case, the system shows an induction period before the crystallization (scenario II). If $r_{w}$ is big enough, the system will never crystallize (scenario III). There is an optimal value of $r_{w}$, denoted as $r_{w}^{0}$, at which the energy provided by the mold exactly matches the interfacial free energy barrier. This value is in the border between scenarios I and II. In $r_{w}^{0}$, $\Delta G^{hw}$ and the solid-fluid interfacial free energy, $\gamma_{hw}$, are related by,

\begin{equation}
\gamma_{hw}=\dfrac{\Delta G^{hw}}{2\mathcal{A}}
\label{eq:interfacial-tension-2}
\end{equation}

\noindent
The factor 2 accounts for the presence of two hydrate-water interfaces when the hydrate slab is formed.

It is essential to emphasize that the initial configuration must be meticulously prepared, ensuring that the system is at coexistence conditions for the valid application of Eq.~\eqref{eq:interfacial-tension-2}. Furthermore, since the interfacial area does not change throughout all the simulations, obtaining an accurate value of $\mathcal{A}$ is crucial to determine $\gamma_{hw}$ appropriately. For further details, we recommend referring to our seminal works.~\cite{Algaba2022b, Zeron2022a, Romero-Guzman2023a}

\section{Simulation details}

 In this work, THF is modeled using a rigid and planar version of the TraPPE-UA model proposed by Keasler~\emph{et al.}~\cite{Keasler2012a} This model has been proposed by some of us in previous papers and it has been demonstrated that the rigid version provides identical results to the original flexible version and reduces significantly the computational effort required by the simulations.~\cite{Garrido2016a,Algaba2018a,Algaba2019a} Water molecule is described using the well-known TIP4P/Ice model.~\cite{Abascal2005b,Conde2017a,Conde2013a} Notice that the choice of the molecular models is far from being arbitrary. This combination has been used previously to determine the dissociation line of the THF hydrate\cite{Algaba2024d}, obtaining an excellent agreement between simulation results and experimental data. Besides, in the original work of Espinosa \emph{et al.},~\cite{Espinosa2016a} the Ih ice--water interfacial tension, $\gamma_{Iw}$, is obtained through the MI technique and using different water models (TIP4P/Ice, TIP4P/2005, TIP4P and mW). They found similar values for $\gamma_{Iw}$ using the TIP4P/Ice (29.8 mJ/m$^2$) and the TIP4P/2005 (28.9 mJ/m$^2$) models. The TIP4P water model provides a value of $\gamma_{Iw}$ slightly lower (27.2 mJ/m$^2$) than the previous ones, and the mW water model provides a higher value (34.9 mJ/m$^2$). Although the four water models provide results of $\gamma_{Iw}$ inside the experimental range (25-35 mJ/m$^2$), the TIP4P/2005 and TIP4P/Ice models provide the most centered values. In addition to that, Conde and Vega\cite{Conde2013a} have shown a correlation between the melting point of the ice I$_{\text{h}}$ and the three-phase coexistence temperature of the methane hydrate determined by several water models. They found that the better the prediction of the melting point of ice I$_{\text{h}}$, the better the prediction of the three-phase coexistence temperature of the methane hydrate. According to Conde and Vega,~\cite{Conde2013a} the two best options to determine accurately the solid phase diagram of water systems (such as ice or hydrate systems) were the TIP4P/Ice and the mW water models. Taking into account the original work of Espinosa \emph{et al.}~\cite{Espinosa2016a} and the work of Conde and Vega,~\cite{Conde2013a} the best option for the hydrate-water $\gamma_{hw}$ calculation is the TIP4P/Ice model. If the model of water is changed, we expect a similar behavior to that found by   Espinosa \emph{et al.}~\cite{Espinosa2016a} in the case of the Ih ice--water interfacial tension, $\gamma_{Iw}$.

 We use a modified Berthelot combining rule to account for the unlike non-bonded dispersive interactions between the oxygen group of water and all the groups of THF. The unlike energy dispersive parameter is given by,
 
 \begin{equation}
 \epsilon_{\text{O}-\text{THF}}=\xi_{\text{O}-\text{THF}}(\epsilon_{\text{OO}}\epsilon_{\text{THF}-\text{THF}})^{1/2}
 \label{eq:combining rules}
 \end{equation}

\noindent
where $\epsilon_{\text{O}-\text{THF}}$ is the well depth of the LJ potential for the unlike interactions between the oxygen of water molecule, O, and THF-groups, $\epsilon_{\text{OO}}$ and $\epsilon_{\text{THF}-\text{THF}}$ are the well depth for the like interactions between O and THF groups respectively and $\xi_{\text{O}-\text{THF}}$ is the factor that modifies the Berthelot combining rule. In this work, we use $\xi_{\text{O}-\text{THF}}=1.4$. This value has been previously proposed by some of us to accurately describe the dissociating or univariant two-phase coexistence line of the THF hydrate, finding an excellent agreement between simulation results and experimental data.~\cite{Algaba2024d} 

We use the GROMACS simulation package~\cite{VanDerSpoel2005a} (version 4.6.5 double-precision) to perform MD simulations via the isothermal-isobaric ensemble~\cite{Allen2017a,Frenkel2002a} at $500\,\text{bar}$ using the MI-H~\cite{Algaba2022b,Romero-Guzman2023a} technique. Notice that the most recent versions of GROMACS don't allow the use of tabulated potentials, which is strictly necessary in order to implement the potential interaction between the molecules of water and the attractive interaction sites of the mold. To calculate the THF hydrate-water interfacial free energy through the MI-H technique, several steps are required: preparation of the simulation box at the thermodynamics conditions at which the interfacial free energy is obtained; determination of the optimal well radius $r_{w}^{0}$; calculation of the interfacial free energy at different values of $r_w>r_w^0$ using thermodynamics integration; and finally, extrapolation of the interfacial free energy to $r_{w}^{0}$. In this section, we focus on the preparation of the simulation box.

The hydrate phase is built by replicating twice the THF hydrate unit cell in the three space directions. The final hydrate simulation box has 1088 water molecules and 64 THF molecules. Note that THF molecules only occupy the tetrakaidecahedron cages, also named as $5^{12}$6$^{4}$, T or large cages, of the sII crystalline structure. This box is equilibrated during $50\,\text{ns}$ at the dissociating conditions of $500\,\text{bar}$ and $272.5\,\text{K}$. The pressure and temperature are fixed using the Parrinello-Rahman isotropic barostat~\cite{Parrinello1981a} and the V-rescale thermostat\cite{Bussi2007a} algorithms respectively. From this simulation, we obtained the average or equilibrium values of $L_{x}$, $L_{y}$, and $L_{z}$, the sizes of the simulation box along the $x$-,  $y$-, and $z$-axis, respectively of the hydrate. Since the sII crystalline structure presents cubic symmetry, the average values $\langle L_{x}\rangle$, $\langle L_{y}\rangle$, and $\langle L_{z}\rangle$ are used to calculate the lattice constant ($a=1.73\,\text{nm}$) at the thermodynamic conditions employed in this work. Then, it is used to build an aqueous simulation box where $L_{x}$ and $L_{y}$ are equal to twice the lattice constant ($3.46\,\text{nm}$) and $L_{z}$ is equal to four times the lattice constant ($6.92\,\text{nm}$). This is congruent with the replication of the unit cell of the sII structure twice along the $x$- and $y$-axis directions, and four times along the $z$-axis direction. This simulation box contains $2176$ and $128$ water and THF molecules in a fluid phase. At this point, it is important to remark that this is the number of molecules expected in a hydrate phase of equal size and, since the aqueous and hydrate phases have the same composition, we can use the same number of molecules to build both phases. Notice that, according to our previous works,~\cite{Blazquez2024a,Algaba2024a,Algaba2024b} the size of the system is large enough to avoid finite-size effects. This box is equilibrated for 50 ns in the ${NP_{z}\mathcal{A}T}$ using the Parrinello-Rahman barostat~\cite{Parrinello1981a} and the V-rescale thermostat\cite{Bussi2007a} algorithms. Along the simulation, $\mathcal{A}=L_{x}\times L_{y}$ remains constant and only fluctuations of volume in $z$ are allowed to equilibrate the pressure. This is crucial since $\mathcal{A}$ represents the size of the equilibrated hydrate as well as the size of the hydrate-water interface.  

\begin{figure*}
\hspace*{-0.7cm}
\centering
\includegraphics[width=2\columnwidth]{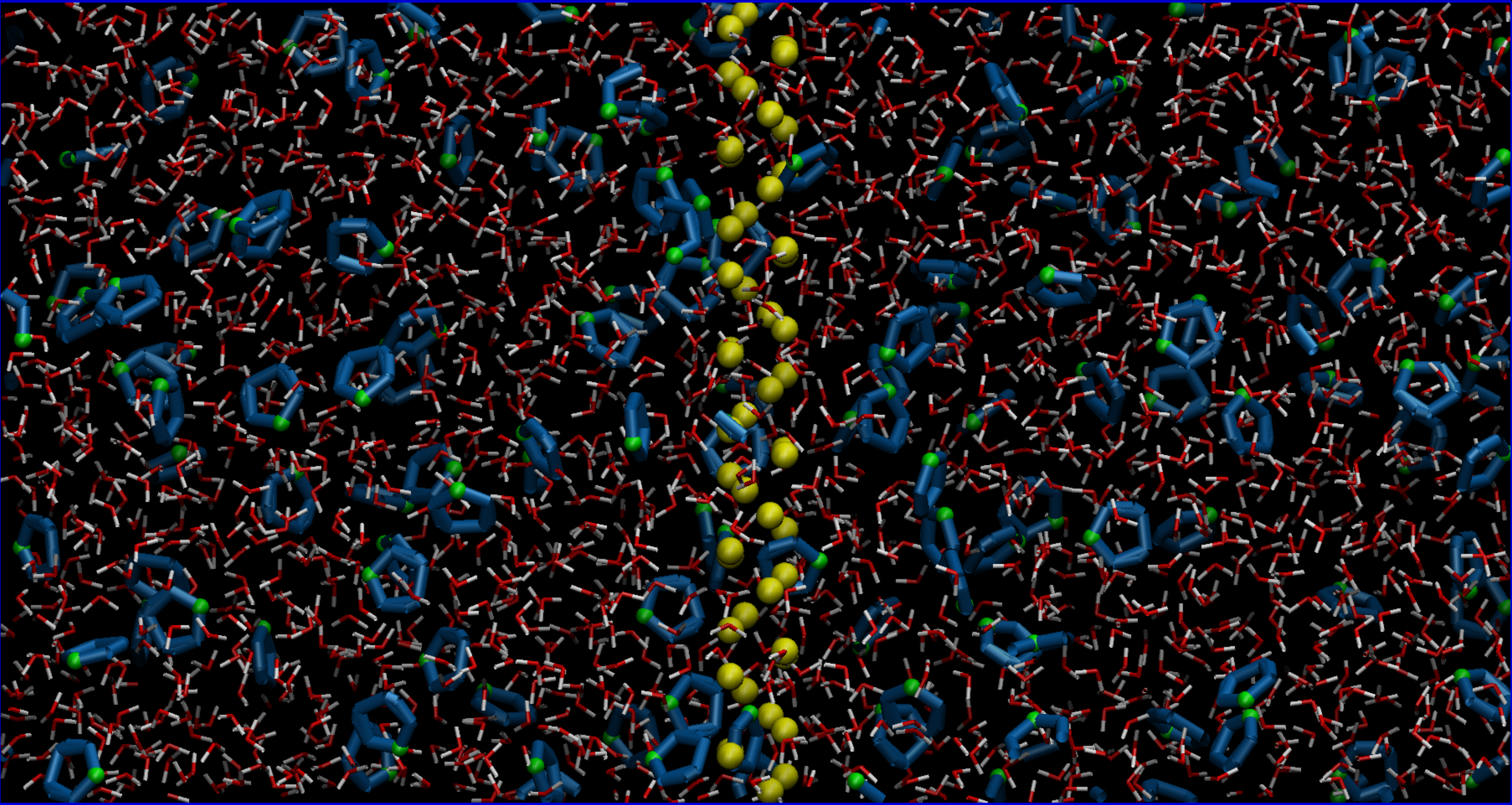}
\vspace*{-0.0cm}
\caption{Snapshot of the initial simulation box, at $500\,\text{bar}$ and $272.5,\text{K}$, used in all the simulations. Water molecules are represented using red (oxygen atoms) and white (hydrogen atoms) licorice representation and THF molecules using cyan licorice representation for the CH$_{2}$ groups and green spheres (van der Waals representation) for oxygen atoms. Yellow spheres (van der Waals representation) in the center of the simulation box represent the attractive interaction sites of the mold.}
\label{figure0}
\end{figure*}

We also place $N_{w}=96$ attractive interaction sites with dispersive energy 
$\varepsilon_{m}=8\,k_{B}T$, in the center of the aqueous simulation box, at the equilibrium positions of the oxygen atoms of the water molecules located at one of the principal planes of the sII crystalline structure. A snapshot of the initial configuration can be visualized in Fig.~\ref{figure0}. Simulations used to determine $r_{w}^{0}$ and to perform the thermodynamic integration are run $100$ and $50\,\text{ns}$, respectively (see Section~IV for further details).

In all cases, the Verlet leapfrog algorithm~\cite{Cuendet2007a} with a time step of $0.002\,\text{ps}$ is used to solve Newton's movement equations. The Parrinello-Rahman barostat~\cite{Parrinello1981a} is used with a time constant of $1\,\text{ps}$ and a compressibility value of $4.5\times 10^{-5}$. The V-rescale thermostat\cite{Bussi2007a} algorithm with a time constant of $0.05\,\text{ps}$ is chosen to fix the temperature value along the simulation. Non-bonded Lennard-Jones and coulombic interactions are truncated using a cut-off distance $r_{c}=1.55\,\text{nm}$. No long-range corrections are applied for the Lennard-Jones interactions and particle-mesh Ewald (PME)~\cite{Essmann1995a} corrections are used for the coulombic potential. Finally, the LINCS algorithm is also used to keep the molecular geometry.

\section{Results}
\begin{figure}
%\hspace*{-0.5cm}
%\includegraphics[width=1.27\columnwidth]{fitted-THF_short.png}
\includegraphics[width=1.0\columnwidth]{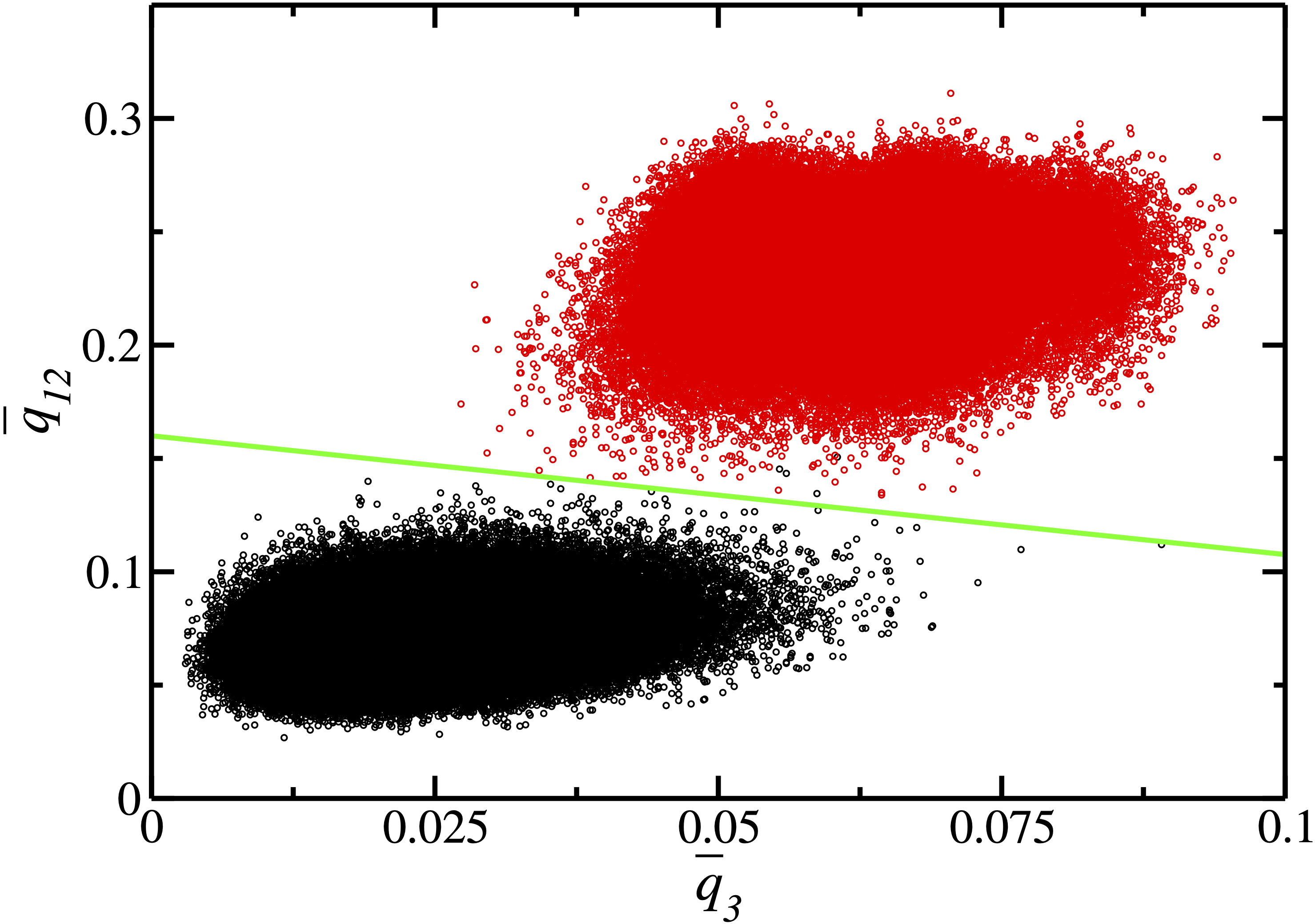}
%\vspace*{-0.8cm}
\caption{Values of $\overline{q}_{12}$ versus $\overline{q}_{3}$, for water molecules at $272.5\,\text{K}$ and $500\,\text{bar}$. The top cloud of points corresponds to water molecules in the THF hydrate phase (red circles) and the bottom cloud to water molecules in the stoichiometric aqueous solution of THF (black circles). The green line represents the linear combination of the order parameters $\overline{q}_{3}$ and $\overline{q}_{12}$ used in this work.}
\label{figure-qs}
\end{figure}

\begin{figure*}[t]
\hspace*{-0.7cm}
\centering
\includegraphics[width=2\columnwidth]{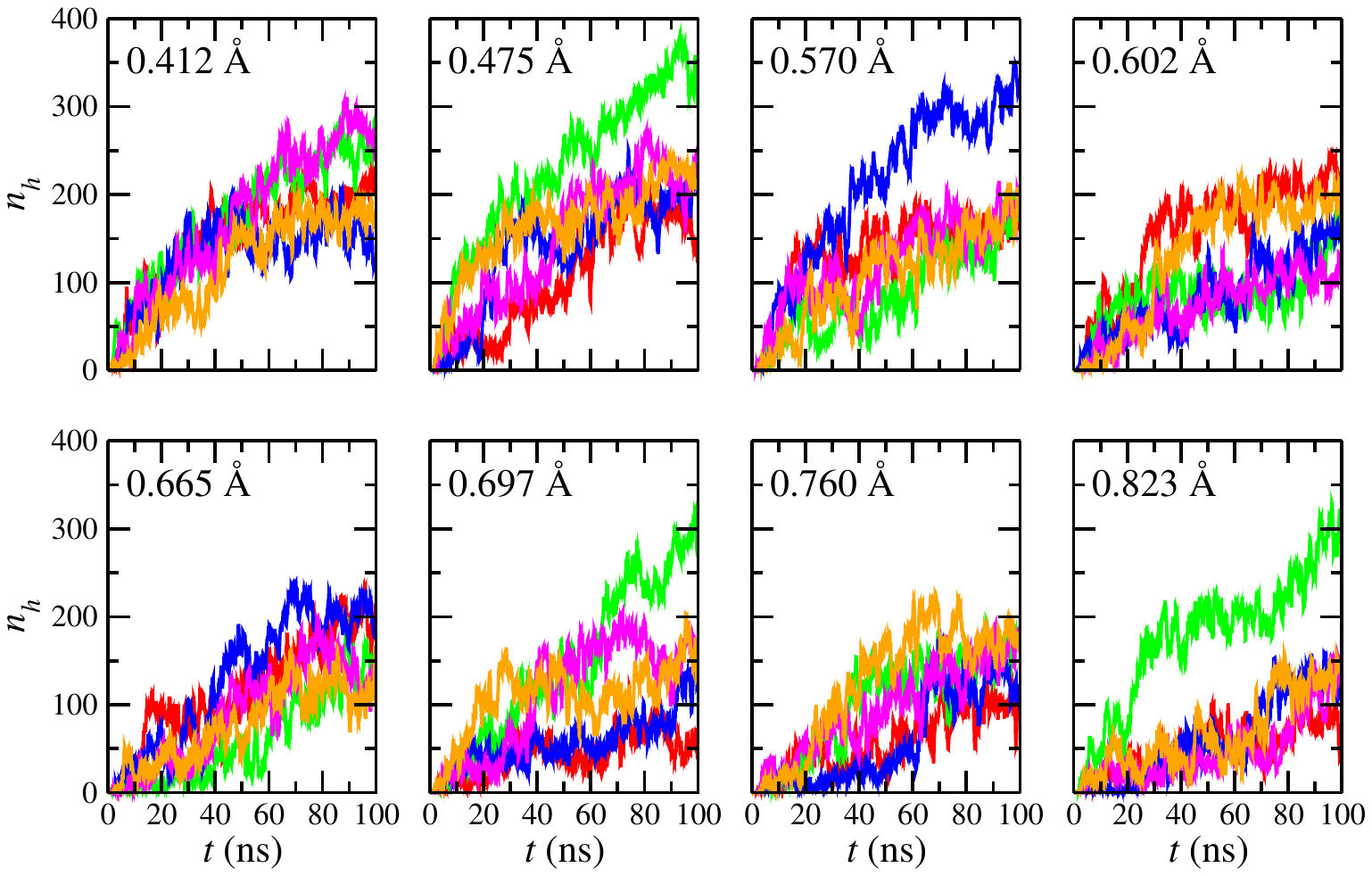}
\vspace*{-0.8cm}
\caption{Number of water molecules in the hydrate slab, $n_{h}$,
 as a function of time, for five trajectories and different well radius $r_{w}$ (as indicated in the legends). Simulations are performed using $\varepsilon=8\,k_{B}T$ and at coexistence conditions ($500 \,\text{bar}$ and $272.5\,\text{K}$). Independent trajectories are represented by different colors, starting from the same initial configuration.}
\label{figure1}
\end{figure*}

As we have already mentioned, it is possible to induce a hydrate slab inside our system by employing the MI-H methodology.~\cite{Algaba2022b,Romero-Guzman2023a} This technique consists of placing a series of potential wells within a fluid under coexistence conditions with its solid. The potential wells are located at the crystallographic positions of the sII structure of the THF hydrate. These potential wells trap the oxygen atoms of water molecules at these positions. By trapping the molecules of water in the mold and forcing them to remain fixed, a solid THF hydrate slab is induced. As it was explained in Sec.~II, the wells provide energy to the system based on their number and depth. The chosen well number and depth in this work are $N_w=96$ and $\varepsilon_{m}=8\,k_{B}T$, respectively.

\subsection{$r_{w}^0$ determination} 

The first step is to find the optimal value of the cut-off well radius, $r_{w}^{0}$. This is the radius value at which the mold provides to the system the exact energy needed to overcome the energetic barrier and crystallize. To study the behavior of the system, as a function of the cut-off well radius, several simulations are performed in the range $r_{w}=0.348-1.267\,\text{\AA}$. The number of water molecules in the hydrate phase, as a function of time, $n_{h}=n_{h}(t)$, is monitored using the local bond order parameters proposed by Lechner and Dellago.~\cite{Lechner2008a} In particular, we propose to use a $\overline{q}_{12}-\overline{q}_{3}$ representation of these averaged order parameters to distinguish between liquid-like and solid-like particles. As can be seen in Fig.~\ref{figure-qs}, this representation allows separating in two different clouds the values obtained from the molecules of water in the hydrate and aqueous phases, allowing to distinguish between liquid-like water molecules (black circles) and hydrate-like water molecules (red circles). As far as the authors know, this is the first time that $\overline{q}_{12}$ and $\overline{q}_{3}$ parameters are used to distinguish between liquid-like and hydrate-liquid molecules in the context of hydrates that exhibit sII crystallographic structure.

Following the standard approach to identify if water molecules are in the liquid or hydrate phase,~\cite{Lechner2008a,Sanz2013a,Espinosa2016c} the $\overline{q}_{12}$ and $\overline{q}_{3}$ parameters are calculated for all the water molecules along the simulations. The green line in Fig.~\ref{figure-qs} can be expressed as a linear combination of the order parameters given by: $\overline{q}_{12}^{nh}=0.16-(0.52407\,\overline{q}_{3})$. If the value of $\overline{q}_{12}$ obtained is higher than $\overline{q}_{12}^{nh}$ (for a particular value of $\overline{q}_{3}$), the molecule of water is labeled as hydrate-like. Contrary, if the value of $\overline{q}_{12}$ obtained is lower than $\overline{q}_{12}^{nh}$, the molecule of water is labeled as liquid-like. The combination $\overline{q}_{12}-\overline{q}_{3}$, shows a better separation of the clouds associated to liquid-like and hydrate-like water molecules than the $\overline{q}_{6}-\overline{q}_{3}$ representation used by some of us in previous works.~\cite{Algaba2022b,Zeron2022a,Romero-Guzman2023a} As can be seen in Fig.~\ref{figure-qs}, this election allows to differentiate liquid-like and hydrate-like water molecules with a mislabeling of $0.003\%$. Finally, to minimize the mislabeling when the molecules of water are labeled as hydrate-like or liquid-like molecules, only the growth of the biggest cluster over the simulation time is analyzed. This strategy has been also used in previous works by other authors.~\cite{Sanz2013a,Espinosa2016c} At this point it is important to remark that the final interfacial free energy value, $\gamma_{hw}$, obtained through the MI methodology is independent of the order parameters chosen. In most simulation techniques, in which the estimation of nucleation rates is obtained through the determination of the size of the critical cluster, the choice of different order parameters affects the final result. However, this is not the case in the MI method where order parameters are only used to determine if the system crystallizes as soon as the simulation begins or if there is an induction period. The final result of $\gamma_{hw}$ and the choice of the optimal well radius is independent of the total number of water molecules that transition from the aqueous to the hydrate phase. As far as the local parameters are able to analyze qualitatively the formation of the hydrate phase, the final result will be the same independently of the local parameters employed.

To identify the optimal value of the cut-off well radius, a minimum of five independent simulations (seeds) are run for each radius value explored. All simulations start from the same initial fluid configuration but with different velocities generated from a random seed. These simulations are run during $100\,\text{ns}$. At this point, our main goal is finding the optimal cut-off well radius, $r_{w}^{0}$, since this is the radius at which $\gamma_{hw}$ is calculated. By representing $n_{h}$, as a function of time, it is possible to distinguish whether the system is crystallizing or not. As can be seen in Fig.~\ref{figure1}, two scenarios are found. From $r_w=0.602$ $\text{\AA}$ to lower $r_w$ values, all simulations crystallize as soon as the simulation begins (scenario I), going through a first-order phase transition upon starting the simulations. At these values of $r_w$, the mold is given to the system more energy than that required to overcome the energy barrier. As a consequence of this, the fluid phase becomes unstable and the system crystallizes immediately after initiating the simulation. From $r_w=0.665$ $\text{\AA}$ to higher $r_w$ values, the simulations exhibit an induction period before crystallizing (scenarios II and III), characterized by an initial flat zone ($n_h\approx0$) before the THF hydrate is formed. At these values of $r_w$, the mold is given to the system less energy than that to overcome the energy barrier. As a consequence of this, the system needs some time before overcoming the energy barrier. Notice that according to the original method,~\cite{Espinosa2014a,Espinosa2016a}
$r_{w}$ is above the optimal value $r_{w}^{0}$ if at least one seed shows induction period. This behavior is clearly shown by the green seed ($r_{w}=0.665\,\text{\AA}$), the blue and red seeds ($r_{w}=0.760\,\text{\AA}$), and the blue and pink seeds ($r_{w}=0.823\,\text{\AA}$). If $r_w\gg r_w^0$, the system will not be able to overcome the energy barrier, and the crystallization will never take place. According to the original MI method,~\cite{Espinosa2014a} $r_{w}^{0}$ is in the middle of the highest radius that does not have an induction period, $r_{w}^{(l)}$, and the smallest radius that shows it, $r_{w}^{(u)}$, or in other words, between scenarios I and II. Taking into consideration this definition and the results shown in Fig \ref{figure1}, the $r_{w}^{0}$ is between $r_{w}^{(l)}=0.602$ $\text{\AA}$ and $r_{w}^{(u)}=0.665$ $\text{\AA}$. At this point, it is interesting to remark that the determination of $r_{w}^{0}$ is the main source of error of the MI method. This is so because sometimes it is difficult to identify without any doubt if the behavior of the system at a given value of $r_w$ corresponds to scenario I or II. That is why it is highly recommended to use a wide range of values of $r_w$ in which it is secured that the $r_{w}^{0}$ is enclosed. For this reason, although Fig. \ref{figure1} shows that $r_{w}^{(l)}=0.602$ $\text{\AA}$ and $r_{w}^{(u)}=0.665$ $\text{\AA}$, we prefer to use $r_{w}^{(l)}=0.570$ $\text{\AA}$ and $r_{w}^{(u)}=0.697$ $\text{\AA}$. According to this, $r_{w}^{0}= (r_{w}^{(l)}+r_{w}^{(u)})/2 = (0.570+0.697)/2\approx 0.6335$ $\text{\AA}$. However, this wider range increases the uncertainty related to the $r_{w}^{0}$ determination that is calculated as follows $\sigma_{r_{w}^{0}}=(r_{w}^{(l)}-r_{w}^{(u)})/2 = 0.0635$ $\text{\AA}$. The final optimal cutoff well radius and its error are expressed as $r_{w}^{0}=0.63(6)$ $\text{\AA}$. Note that a wider range only affects the uncertainty of $r_{w}^{0}$, and the THF hydrate-water interfacial free energy, but it does not affect the final $r_{w}^{0}$ value.

\subsection{Thermodynamic integration}

Despite knowing $r_{w}^{0}$, it is not possible to perform the thermodynamic integration at this value because the energy given by the potential wells exactly matches the one needed for the system to overcome the energetic barrier and crystallize. This means that the slightest thermal fluctuation can provoke the system crystallization, crossing a first-order phase transition and pushing the system away from its equilibrium state, where Eqs.~\eqref{eq:thermo-int-1} and \eqref{eq:interfacial-tension-2} cannot be applied. Note that $\Delta G^{hw}$ is the reversible work required to form a hydrate slab in the aqueous phase. To avoid this issue, it is necessary to perform the thermodynamic integration at higher radius values where the crystallization never takes place (scenario III). In this work, scenario III arises when $r_w\ge1.203$ $\text{\AA}$.  For these values of $r_w$, the system seems to exhibit an infinite induction period in which no phase transition occurs. By carrying out simulations with $r_w\ge1.203$ $\text{\AA}$, the energy provided by the wells of the mold is not enough to overcome the crystallization energy barrier. This ensures the reversibility of the system, a necessary condition to apply Eqs.~\eqref{eq:thermo-int-1} and \eqref{eq:interfacial-tension-2}. As we have pointed out previously, it is necessary to perform the thermodynamic integration for high $r_w$ values where the crystallization never takes place. In order to calculate accurately the reversible work required to form a hydrate slab in the aqueous phase ($\Delta G^{hw}$), the thermodynamic integration is only applicable under reversibility conditions. For $r_w<1.203\,\text{\AA}$ values, the system can overcome the energy barrier and crystallize after a certain time, going through a first-order phase transition and providing an erroneous estimation value of $\gamma_{hw}$. At this point, one could think a priori that any $r_w$ value higher than $1.203\,\text{\AA}$ can be used to perform the thermodynamic integration. However, following the MI methodology, only one oxygen atom of water must occupy each well. This condition is not met if $r_w$ and $\varepsilon_m$ are large enough. According to this, large values of $r_w$ can not be chosen in order to avoid double occupancy of the mold wells. In this work, the thermodynamic integration is carried out using the following values for the cut-off well radius: $r_{w}=1.203$, 1.235, and 1.267 $\text{\AA}$. These values ensure that the crystallization of the system and the double occupancy of the mold wells never take place.

It is important to remark that the MI methodology has been applied to determine the solid-fluid interfacial free energy of different systems, including simple molecules that interact via spherical hard-spheres and Lennard Jones intermolecular potentials,~\cite{Espinosa2014a} and more complex molecular systems such as water~\cite{Espinosa2016a} and NaCl.~\cite{Espinosa2015a} The technique has also been used to estimate hydrate-water interfacial free energies of CO$_{2}$ hydrates at different thermodynamic conditions.~\cite{Algaba2022b,Zeron2022a,Romero-Guzman2023a} Despite the different systems and conditions at which the methodology has been applied, interfacial energy has always shown a linear behavior with the $r_{w}$ value. This allows to calculate accurately $\gamma_{hw}$ at the $r_w$ values where the thermodynamic integration is performed (i.e. where the crystallization of the system never takes place) and extrapolate the $\gamma_{hw}$ value at $r_w^0$. In addition to this, agreement between theoretical predictions and experimental data taken from the literature is excellent in all cases.

\begin{figure}
\hspace*{-0.3cm}
\centering
\includegraphics[width=1.25\columnwidth]{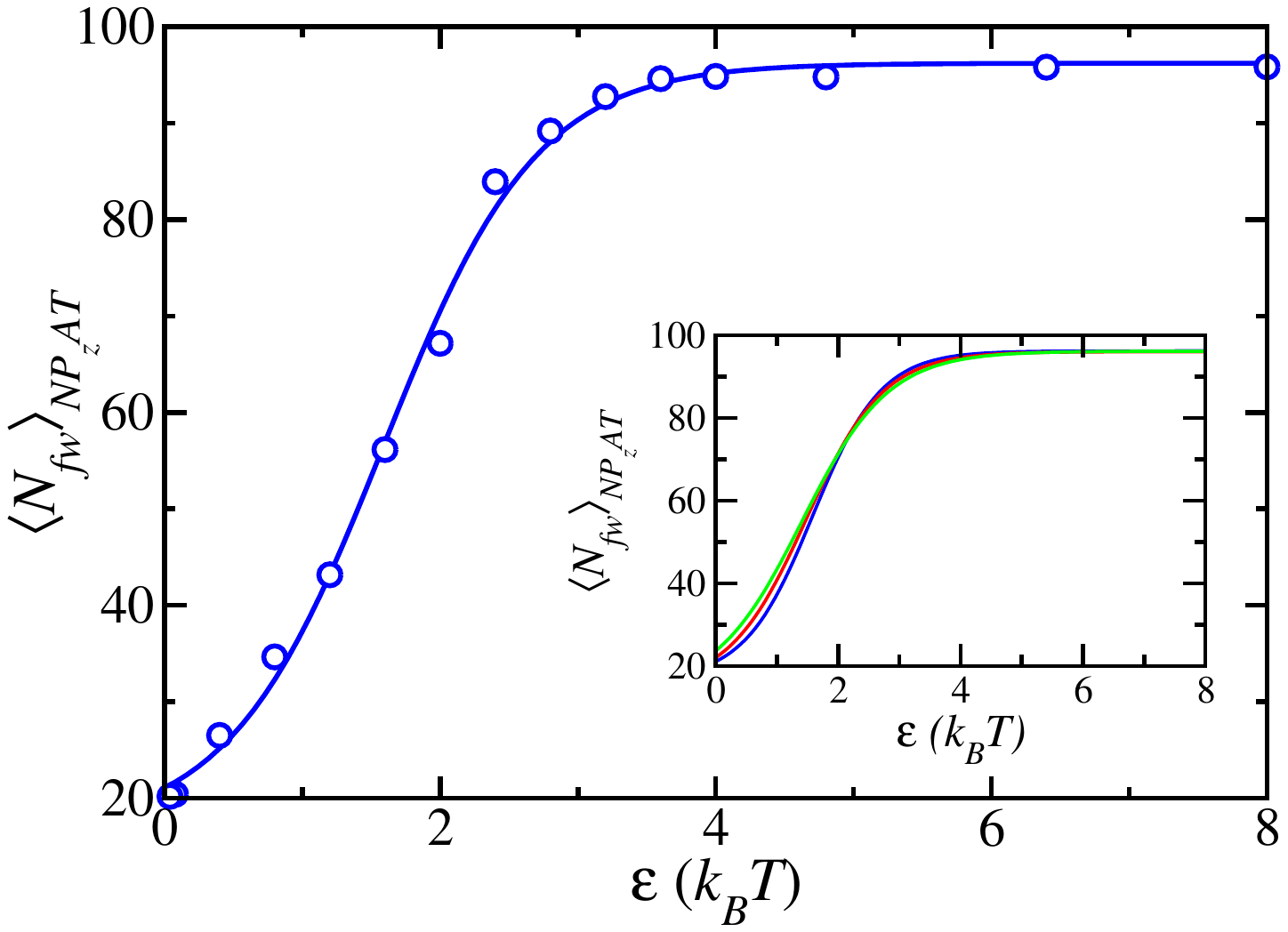}
\vspace*{-0.8cm}
\caption{Average number of filled wells, $\langle N_{fw}\rangle _{NP_{z}\mathcal{A}T}$, as function of the well depth, $\varepsilon$, at $500\,\text{bar}$ and $272.5\,\text{K}$ using a well radius of the mold $r_{w}=1.203\,\text{\AA}$. The blue circles correspond to the values obtained from $NP_{z}T\mathcal{A}$ simulations and the blue curve has been obtained by fitting the results to a hyperbolic tangent function. The inset represents the fits of the average number of filled wells using the well radii of 1.203 (blue), 1.235 (red), and $1.267\,\text{\AA}$ (green).}
\label{figure2}
\end{figure}

To perform the thermodynamic integration it is necessary to evaluate the average number of wells of the mold that are filled by water molecules, $\langle N_{fw}\rangle _{NP_{z}\mathcal{A}T}$, as a function of the value of the attractive interaction potential between the mold and the molecules of water. The attractive interactions go from $\varepsilon=0$ to $\varepsilon_m$. To this end, we follow previous approaches~\cite{Espinosa2014a,Espinosa2016a,Algaba2022b,Zeron2022a,Romero-Guzman2023a} and perform $15$ simulations with different values of $\varepsilon$ for each thermodynamic integration. For each $\varepsilon$ value, the system is first equilibrated during $10\,\text{ns}$ and then the number of wells occupied by water molecules is averaged for another $40\,\text{ns}$. This allows us to obtain a curve that exhibits a smooth behavior of 
$\langle N_{fw}\rangle_{NP_{z}\mathcal{A}T}$, as a function of $\varepsilon$, to perform safely the thermodynamic integration. The curves obtained in this work for different $r_{w}$ values are shown in Fig.~\ref{figure2}.  Particularly, the election of the $r_{w}$ values ensures that there are neither discontinuities nor abrupt leaps in the curves. When the mold is completely activated ($\varepsilon=\varepsilon_m$), the integration curve reaches a plateau and $\langle N_{fw}\rangle_{NP_{z}\mathcal{A}T}\rightarrow N_w$, as can be clearly seen in Fig.~\ref{figure2}. This is one of the conditions that have to be fulfilled to apply the MI-H method: when the mold is completely activated, the molecules of water cannot escape from the mold and, as a consequence, the average number of wells filled by a molecule of water has to be equal to the total number of wells in the mold, $N_w$ ($96$ in this particular case). Finally, we have checked that no more than one water molecule could be found inside a potential well. This is more likely to happen using large radii, conditions at which the thermodynamic integration is carried out. To this end, we monitor the average number of wells occupied by water molecules for each value of $\varepsilon$ and ensure that this number is always lower or equal to the total number of wells in the mold, $N_{w}$.

\subsection{Interfacial free energy}

Once the average of the number of wells that are filled by water molecules is calculated at each radius value and at different values of $\varepsilon$, $\Delta G^{hw}$ can be easily calculated using Eq.~\eqref{eq:thermo-int-1} by integrating the resulting curves at each radius shown in Fig.~\ref{figure2}. $\Delta G^{hw}$ represents the reversible work needed to form a hydrate slab inside the fluid under coexistence conditions. At this point, it is important to remark that the energy given by the mold to the system is subtracted, which means that $\Delta G^{hw}$ is independent of the values of $N_w$ and $\varepsilon_m$ chosen. Finally, and taking into account that the THF hydrate-water interfacial area is $\mathcal{A}=11.99\,\text{nm}^{2}$, $\gamma$ is calculated straightforwardly from Eq.~\eqref{eq:interfacial-tension-2}.

\begin{figure}
\hspace*{-0.4cm}
\centering
\includegraphics[width=1.2\columnwidth]{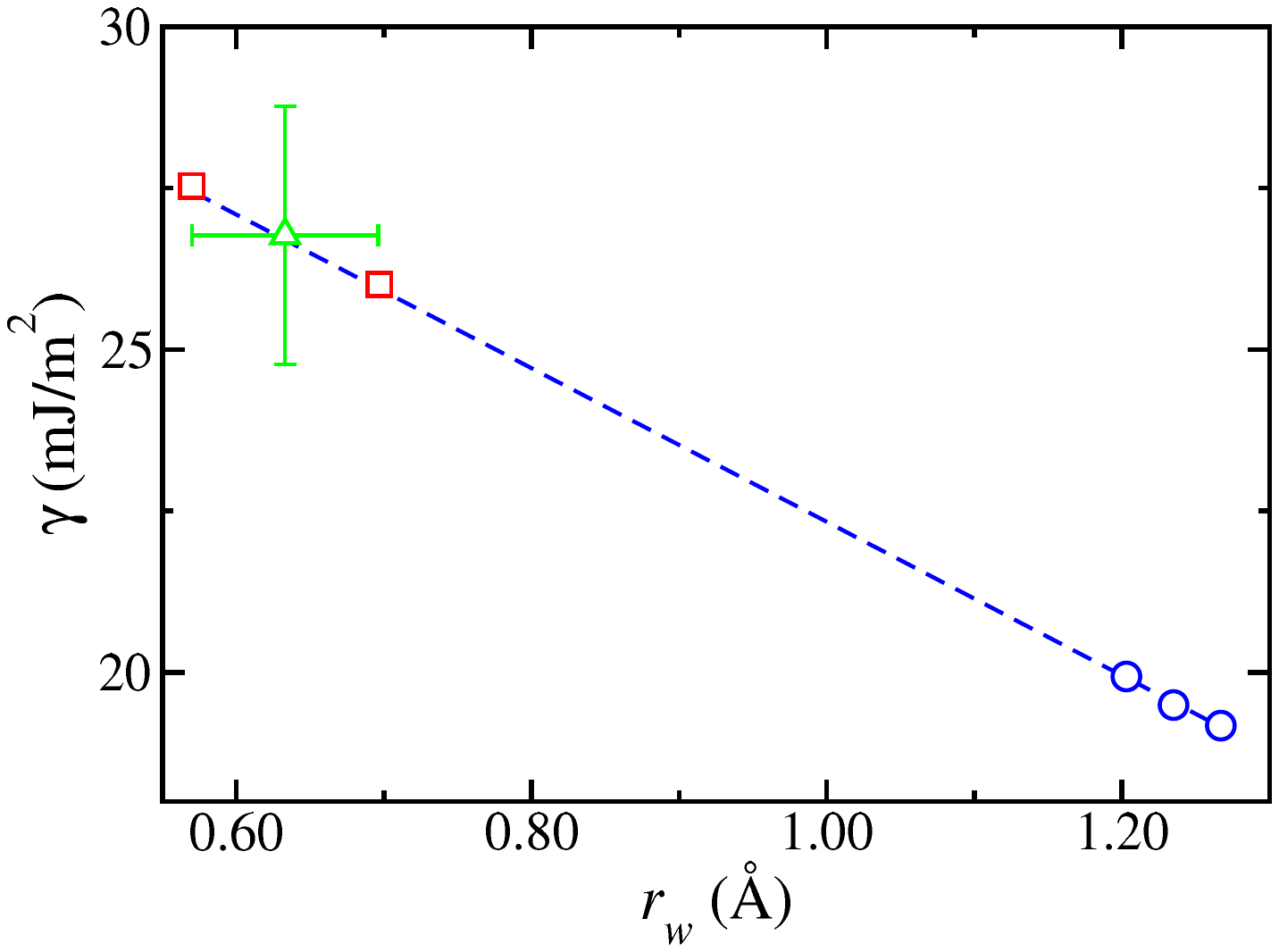}
\vspace*{-0.8cm}
\caption{THF hydrate-water interfacial free energy, as function of the well radius, at $500\,\text{bar}$ and $272.5\,\text{K}$ (open blue circles). The dashed blue line represents a linear fit of the data, the open red squares the interfacial tension evaluated at $r_{w}^{(l)}$ and $r_{w}^{(u)}$, and the open green triangle the extrapolation of the linear fit to the optimal well radius.}
\label{figure3}
\end{figure}

The resulting interfacial free energies obtained for $r_w=1.203$, $1.235$, and $1.267\,\text{\AA}$ are $19.93$, $19.50$, and $19.18\,\text{mJ/m}^{2}$, respectively. Fig.~\ref{figure3} shows the THF hydrate-water interfacial free energy, $\gamma$, as a function of $r_{w}$. As can be seen, the results follow a linear fit in agreement with previous results.~\cite{Espinosa2014a,Espinosa2016a,Algaba2022b,Zeron2022a,Romero-Guzman2023a} According to this, it is possible to extrapolate the $\gamma$ when $r_w\rightarrow r_{w}^{0}$ and obtain $\gamma_{hw}\approx 26.77\,\text{mJ/m}^{2}$. It is also possible to provide an estimation of the uncertainty associated with the interfacial free energy, $\sigma_{\gamma_{hw}}$. Following previous works,~\cite{Espinosa2014a,Espinosa2016a,Algaba2022b,Zeron2022a,Romero-Guzman2023a} $\gamma_{hw}$ uncertainty is determined as half of the difference between the interfacial energies at $r_{w}^{(l)}$ and $r_{w}^{(u)}$, which are 27.53 and 26.01 mJ/m$^2$, respectively. The $\sigma_{\gamma_{hw}}$ value obtained with this approach is $\approx$1 mJ/m$^2$. However, since the main source of error from the MI-H method comes from the election of $r_w^0$, it is highly recommended to provide a generous error bar value. In this work, we estimate $\sigma_{\gamma_{hw}}\approx 2\,\text{mJ/m}^{2}$. According to this, the THF hydrate-water interfacial free energy at $500\,\text{bar}$ and $272.5\,\text{K}$ is $\gamma_{hw}=27(2)\,\text{mJ/m}^{2}$. Figure \ref{figure3} summarizes the results of $\gamma$ at each radius, the extrapolation line, the values of $\gamma$ at $r_{w}^{(l)}$ and $r_{w}^{(u)}$, as well as the final value of $\gamma_{hw}$, at $r_w^0$, with the corresponding uncertainties.

As far as the authors know, there is only an experimental value of the THF hydrate-water interfacial free energy reported by Lee \emph{et al.}~\cite{Lee2007a} In this work, the authors compile experimental data on the properties of THF, methane, their hydrates, and water ice from different authors. According to Table I of that work, the THF hydrate-water interfacial free energy, $\gamma_{hw}$, is between $16$ and $31\,\text{mJ/m}^{2}$. This value of $\gamma_{hw}$ was estimated from the study of Zakrzewski and Handa~\cite{Zakrzewski1993a} of the formation of the THF hydrate in Vycor glass pores of approximated fixed radius pores. Particularly, these authors measured the melting temperature of the pore hydrate and the enthalpy of melting in confined conditions. Knowing the amount depressed of these magnitudes relative to the corresponding values for the bulk phase, it is possible to estimate the interfacial free energy using the well-known Gibbs-Thomson relationship.~\cite{Handa1992a,Clennell1999a,Henry1999a} Although it is relatively easy to measure experimentally fluid-fluid interfacial energies, it is well known that there is no general methodology for determining solid-fluid interfacial energies. In fact, it is accepted that solid-fluid interfacial energies are difficult to measure. The same is true for estimating interfacial energies from computer simulation. Contrary to the MI methodology where the interfacial tension value is calculated at a certain thermodynamic condition of $T$ and $P$, the (indirect) experimental value of $\gamma_{hw}$ calculated by the Gibbs-Thomson relationship is obtained assuming, among other approximations, that the interfacial free energy is constant and it only depends on the mean diameter of the pores. According to this, the hydrate interfacial free energy does not vary with the pressure and it remains constant along the whole hydrate dissociation line. In other words, the dependence of $\gamma_{hw}$ with pressure is not known experimentally and, a priori, the THF hydrate-water interfacial tension calculated at any point of the dissociation line can be compared directly with the only available experimental point.~\cite{Lee2007a}

It is possible to express the experimental value and its uncertainty using the lower and upper values of $\gamma_{hw}$ reported by Lee \emph{et al.},~\cite{Lee2007a} similarly to the approach presented in this Section. According to this, the experimental value can be expressed as $\gamma_{hw}\approx 23.5(7.5)\,\text{mJ/m}^2$. Considering only the most significant figures for the value and the uncertainty and rounding off appropriately, the final value of the interfacial free energy and its uncertainty is $\gamma_{hw}=24(8)\,\text{mJ/m}^{2}$. The first conclusion is that the agreement between the THF hydrate-water interfacial free energy obtained in this work, $\gamma_{hw}=27(2)\,\text{mJ/m}^{2}$ and the experimental data proposed by Lee and coworkers~\cite{Lee2007a} is excellent. To the best of our knowledge, this is the first time that the THF hydrate-water interfacial free energy is calculated using computer simulation for a sII crystalline structure. This result demonstrates that the MI-H methodology, one of the extensions of the original Mold Integration technique, proposed to deal with hydrate-water interfacial free energies, can be used with confidence in hydrates that exhibit a sII structure. This will help to expand the knowledge about the thermodynamics of one of the most used and interesting hydrates, the THF hydrate.

\section{Conclusions}

We have determined the THF hydrate-water interfacial free energy from MD computer simulations. Particularly, we have used an extension of the original Mold Integration technique,~\cite{Espinosa2014a} the Mold Integration-Host (MIH) methodology recently proposed by some of us~\cite{Algaba2022b} to deal with interfacial free energies of clathrate hydrates. According to the method, the calculations must be performed at thermodynamic conditions where the hydrate and fluid phases coexist. To this end, we have used a rigid version of the TraPPE-UA model to describe THF molecules~\cite{Garrido2016a,Algaba2018a,Algaba2019a} and the well-known TIP4P/Ice model for water molecules.~\cite{Abascal2005b} We have recently used the combination of these two models to accurately describe the univariant two-phase dissociation line of the THF hydrate, in a wide range of pressures, from computer simulation.~\cite{Algaba2024d} In this work, we concentrate on the interfacial free energy at $500\,\text{bar}$ at the dissociation line of the THF hydrate.

One of the key points of the MI-H methodology is the use of order parameters to monitor the induced growth of a thin slab of hydrate in the aqueous solution phase of THF. Following our previous works, we use the Lechner and Dellago bond local order parameters~\cite{Lechner2008a} to identify if water molecules are hydrate-like or liquid-like. We have discovered that the combination of the  $\overline{q}_{12}$ and $\overline{q}_{3}$ averaged local order parameters, used to successfully distinguish between hydrate-like and liquid-like water molecules in hydrates that exhibit sI crystalline structure,~\cite{Grabowska2022b} can also be used with confidence in the context of sII hydrates.

The THF hydrate-water interfacial free energy predictions obtained in this work are compared with the only indirect experimental value existing in the literature estimated by Lee and coworkers (via the Gibbs-Thomson equation)~\cite{Lee2007a} from experimental data measured by Zakrzewski \emph{et al.}~\cite{Zakrzewski1993a} The value of the interfacial free energy obtained in this work, $27(2)\,\text{mJ/m}^{2}$, is in excellent agreement with the experimental data, $24(8)\,\text{mJ/m}^{2}$. To the best of our knowledge, this is the first time that the THF hydrate-water interfacial free energy is calculated through computer simulation.

This work confirms that the original Mold integration technique,~\cite{Espinosa2014a} including the two extensions to deal with interfacial free energies of hydrates,~\cite{Algaba2022b,Zeron2022a} can be used with confidence to predict solid-fluid interfacial free energies of simple models, including the hard-sphere and Lennard-Jones systems,~\cite{Espinosa2014a} water,~\cite{Espinosa2016b} and salt aqueous solutions,~\cite{Soria2018a,Sanchez-Burgos2023a} but also more complex crystalline solid phase, including structures sI and sII exhibited by CO$_{2}$~\cite{Algaba2022b,Zeron2022a,Romero-Guzman2023a} and THF hydrates.

\section*{Acknowledgements}

This work was funded by Ministerio de Ciencia e Innovaci\'on (Grant No.~PID2021-125081NB-I00), Junta de Andalucía (P20-00363), and Universidad de Huelva (P.O. FEDER EPIT1282023), all three co-financed by EU FEDER funds. CR-G acknowledges the FPI Grant (Ref.~PRE2022-104950) from Ministerio de Ciencia e Innovaci\'on and Fondo Social Europeo Plus. MJT also acknowledges the research contract (Ref.~01/2022/38143) of Programa Investigo (Plan de Recuperaci\'on, Transformaci\'on y Resiliencia, Fondos NextGeneration EU) from Junta de Andaluc\'{\i}a (HU/INV/0004/2022). We greatly acknowledge RES resources at Picasso provided by The Supercomputing and Bioinnovation Center of the University of Malaga to FI-2024-1-0017.

\section*{Author declarations}

\noindent
\textbf{Conflict of interests}

The authors declare no conflicts to disclose.

\section*{Author contributions}

\noindent
\textbf{Miguel J. Torrej\'on:} Visualization (lead); Methodology (equal); Investigation (lead); Writing – review \& editing (equal).
\textbf{Crist\'obal Romero-Guzm\'an:} Visualization (lead); Methodology (equal); Investigation (lead); Writing – review \& editing (equal).
\textbf{Manuel M. Piñeiro:} Funding acquisition (lead); Methodology (equal); Writing – original draft (lead); Writing – review \& editing (equal).
\textbf{Felipe J. Blas:} Conceptualization (lead); Funding acquisition (lead); Methodology (equal); Writing – original draft (lead); Writing – review \& editing (equal).
\textbf{Jesús Algaba:} Conceptualization (lead); Methodology (equal); Writing – original draft (lead); Writing – review \& editing (equal).

\section*{Data availability}

The data that support the findings of this study are available within the article.

\section*{REFERENCES}
%\nocite{*}
\bibliography{bibfjblas}% Produces the bibliography via BibTeX.

\end{document}